\documentclass[manuscript]{aastex}

\received{}
\revised{}
\accepted{}

\shortauthors{Jones {\it et al.}}
\shorttitle{NGC 3628}

\begin{document}

\title{The 3D Morphology of VY Canis Majoris II: Polarimetry and the Line-of-Sight Distribution of the Ejecta~\altaffilmark{1}}
\author{Terry Jay Jones, Roberta M. Humphreys, and L. Andrew Helton}
\affil{Department of Astronomy, University of Minnesota, 116 Church 
Street S.E., \\ Minneapolis, MN 55455 \\ tjj@astro.umn.edu, roberta@aps.umn.edu, ahelton@astro.umn.edu}
\author{Changfeng Gui~\altaffilmark{2} and Xiang Huang}
\affil{Department of Mathematics, University of Connecticut, 196 Auditorium Road University of Connecticut, U-3009 Storrs, CT 06269-3009 \\ gui@math.uconn.edu, huang@math.uconn.edu}

\altaffiltext{1}{Based on observations with the NASA/ESA Hubble Space Telescope obtained at the Space Telescope Science Institute, which is operated by the Association of Universities for Research in Astronomy, Inc. under NASA contract NAS 5-26555.}

\altaffiltext{2}{Visitor, Institute for Mathematics and its Applications, University of Minnesota, 127 Vincent Hall, 206 Church St. S.E., Minneapolis, MN 55455} 

\begin{abstract}

We use imaging polarimetry taken with the HST/ACS/HRC to explore the three dimensional structure of the circumstellar dust distribution around the red supergiant VY Canis Majoris. The polarization vectors of the nebulosity surrounding VY CMa show a strong centro-symmetric pattern in all directions except directly East and range from 10\% - 80\% in fractional polarization. In regions that are optically thin, and therefore likely have only single scattering, we use the fractional polarization and photometric color to locate the physical position of the dust along the line-of-sight. Most of the individual arc-like features and clumps seen in the intensity image are also features in the fractioanl polarization map. These features must be distinct geometric objects. If they were just local density enhancements, the fractional polarization would not change so abruptly at the edge of the feature. The location of these features in the ejecta of VY CMa using polarimetry provides a determination of their 3D geometry independent of, but in close agreement with, the results from our study of their kinematics (Paper I).

\end{abstract}

\keywords{stars: mass loss --- stars: supergiants -- stars: individual (VY CMa) } 

\section{Introduction} \label{sec:intro}

VY CMa is one of the most luminous evolved stars known, and is near the empirical upper limit for cool hypergiants \citep{HD, dej98}. At a distance of 1.5 kpc, \citep{her72, lad78, mar97}, the star has a luminosity of $\sim 4.3\times 10^{5}L_{\sun}$. VY CMa is surrounded by an asymmetric nebula, 10'' across, created by a very high mass loss rate of $4 \times 10^{-4}M_{\sun} yr^{-1}$ \citep{dan94}. This high mass loss rate makes VY CMa a very important star for studying the process by which evolved cool stars return material to the interstellar medium.

Multi-wavelength HST/WFPC2 images of VY CMa \citep{smi01} revealed a complex circumstellar environment dominated by multiple filamentary arcs and knots. The apparent random orientations of the arcs suggested that they were produced by localized ejections, not necessarily aligned with either the star's presumed NE/SW axis \citep{mor80, bow83, ric98} or its equator. Therefore, Smith {\it {et al.}} speculated that the arcs may be expanding loops caused by localized activity on the star's ill-defined surface. To learn more about the morphology, kinematics and origin of VY CMa's complex ejecta, \citet{hum05} obtained long-slit spectra with HIRES on the Keck 1 telescope to map the emission and absorption lines in the nebula. They found that VY CMa shows evidence not only for multiple and asymmetric mass loss events at different times, but also ejections that are recognizably and kinematically separate from the general flow of the diffuse material.

Radial velocities alone can not reveal the overall expansion of the nebula, the direction of the outflows, and orientation of the arcs within the surrounding nebulosity. Fortunately, VY CMa provides us with a unique opportunity to determine the three dimensional morphology of its ejecta and the geometry of the discrete structures embedded in it. In \citet[hereafter Paper I] {hum07}, we analyzed second epoch WFPC2 images and were able to measure the transverse motions of several features in the nebula. These proper motions were combined with the radial velocity measurements for 40 different positions to produce full space motions for the prominent arcs and knots in the nebulosity. Under the assumption that these space motions were purely radial with respect to the star, the arcs and knots could be located in three dimensional space with respect to the central star. 

In this paper (hereafter Paper II) we employ a technique utilizing polarimetry as an independent method for studying the 3D geometry of the ejecta surrounding VY CMa. Like other reflection nebulae, VY CMa is highly polarized. \citet{her72} measured polarization up to 70\% in the nebula, but no polarimetry has been done on VY CMa since his groundbased photographic measurements. We have obtained polarimetric images with the Advanced Camera for Surveys High Resolution Camera (ACS/HRC) on the HST with sufficient resolution to resolve all of the features seen in our previous work, but in fractional polarization. The observed fractional polariation, combined with our photometric color images \citep{smi01}, can be used to compute a scattering angle for the dust feature. Knowledge of the scattering angle and the angular position of the dust from the star (on the plane of the sky) allows us to place the scattering material in three dimensions with respect to the star.

\section{Observations and Data Reduction} \label{sec:obs}

VY CMa was observed on 2004 August 17 using the Advanced Camera for Surveys (ACS) High Resolution Camera (HRC) which has a $29 \times 25 \arcsec$ field of view with a plate scale of $\sim 0.027 \arcsec$/pixel. Images were taken using three polarizing filters, Pol0V, Pol60V and Pol120V through both the F550M medium width visual filter and the F658N $H\alpha$ narrow band filter.  The images obtained with the F550M filter were single exposures at 0.2, 0.5, 5.0 and 60 seconds while the $H\alpha$a images were single exposures at 1.0, 5.0, 40 and 150 seconds.  All images were processed using the CALACS pipeline version 4.5.1 which performs standard calibration procedures including bias and dark subtraction and flat fielding.  This was followed by MultiDrizzle for distortion correction and flux conversion.  Since there was only one exposure per integration time per filter combination, cosmic ray rejection was performed by hand using the IRAF task FIXPIX to interpolate over cosmic rays and bad pixels that were not removed by MultiDrizzle. Images were aligned using IRAF procedures IMCENTROID and IMALIGN.

We have used the standard formulae to compute the polarization from the three polarized images.

\begin{eqnarray*}
 I  &=& \frac{2}{3}\left( {I_0  + I_{60}  + I_{120} } \right) \\ 
 Q &=& \frac{2}{3}\left( {2I_0  - I_{60}  - I_{120} } \right) \\ 
 U &=& \frac{2}{{\sqrt 3 }}\left( {I_{60}  - I_{120} } \right) \\ 
\end{eqnarray*}
where we also define the instrumental fractional polarization vector quantities:
\begin{eqnarray*}
 P_X  = Q/I \\ 
 P_Y  = U/I \\ 
\end{eqnarray*}

Forming Stokes Q and U images requires very accurate relative positioning of the three images taken through the polarizers. Changes in X and Y shifts between images, even as small as a fraction of a pixel, can produce spurious Q and U signals. For example, a small differential shift between two otherwise identical images of a star will produce positive and negative signals when the images are subtracted. We have used techniques we developed for ground based imaging polarimetry of comets and galaxies \citep{jon00, jon97} to form Q and U intensity images with minimal spurious Q and U signal.

Although the HST has a polarimetric capability, it is not a precision polarimeter and suffers from complicated instrumental polarization effects. We need to consider a number of sources of error in our polarimetry. One is the overall level of instrumental polarization, the measured polarization for an unpolarized source (`diattenuation' in the ACS Data Handbook, also commonly known as differential transmission). Another is the `cross polarization', the mixing of Q, U and V Stokes parameters (`phase retardance' in the ACS Handbook, also known as rotation). A third consideration is the variation in instrumental effects across the HRC field. Several authors have worked on these issues with specific configurations of the HST, none of which are exactly the same as ours \citep[e.g.][]{sch99, per06}. Typical results from efforts to characterize HST instrumental polarization yield +/- 3\% as the level of instrumental polarization and factors of +/1 0.1 (10\%) mixing of Q/I and U/I. \citet{koz05} have mapped out the variations in the instrumental response through several combinations of filters and polarizers across the ACS field. These variations are modest, a few percent or two over the field. These results are not available in numerical form and since we are dealing with a high overall polarization in the VY CMa nebula, we will not make any spatial corrections. 

We can, however, estimate the combined effect of the instrumental polarization and interstellar polarization for our particular HST configuration and observing in the direction of VY CMa, using a nearby star, GSC6541-02318. There are no previous polarimetry observations of this star, so we do not know if it has any significant intrinsic or interstellar polarization. Figure 1 shows the measured interstellar polarization of stars in the vicinity of VY CMa taken from the compilation of \citet{hei00}. Interstellar polarization is always less than 0.5\%, an insignificant amount for our purposes. The stars in the compilation of Heiles are realtively nearby, however, compared to VY CMa and possibly GSC6541-02318. We will assume GSC6541-02318 is intrinsically unpolarized and use our measurements of its polarization with the HST to determine a `total' instrumental + interstellar polarization to be subtracted from the observations of VY CMa.

GSC6541-02318 was placed in three different locations on the HRC CCD, the center and two locations $5\arcsec$ off center. The measured values for these locations in the F658N filter are given in Table 1 in terms of the instrumental $P_X$ and $P_Y$. The mean instrumental polarization is 6.6\% with variations of less than 1\%. The magnitude of the measured polarization is consistent with pure instrumental polarization if we refer to Figure 6.8 in the ACS Data Handbook. Table 6.7 in the ACS Data Handbook lists correction factors for the Pol0V, Pol60V, and Po1120V polarizing filters in the F658N filter to be 1.061, 0.971 and 0.973 respectively. Our average values for GSC6541-02318 were 1.058, 0.971 and 0.973, in excellent agreement under the assumption that GSC6541-02318 is unpolarized and we are measuring only instrumental polarization. We have corrected the entire field using this value for the instrumental polarization, ignoring spatial variations. We estimate errors in the removal of both instrumental polarization and variations in the instrumental polarization across the field to be 1-3\%.  Given the high polarization in the VY CMa nebula, 10-80\%, these uncertainties will not significantly impact our analysis.

Finally, we ignore any possible effects of cross polarization, for which we do not have sufficient data to accurately measure across the field. The ACS Data Handbook suggests these effects in polarization magnitude are at the one-part-in-ten level with an uncertainly of about $3\degr$ in position angle.  We will only be considering differences in polarization of 10\% or greater in our analysis, and variations in position angle of $3\degr$ are irrelevent.

The resulting polarization vector map of the inner portions of the nebula is shown in Figure 2. Our high resolution results can be compared with the photographic polarization map from \citet{her72} that covers a larger area on the sky (mostly external to the small region imaged by the HST). The overall centro-symmetric scattering pattern seen in Herbig's polarimetry is present in our HST polarimetry map. We note, however, that the fainter region directly East and and SE has a low polarization and shows significant deviation from centro-symmetric scattering in the polarization position angles. This is also true in Herbig's map $2-4\arcsec$ directly to the East of the star. We also note there are small, residual effects due to the spider arms in the polarization map.

A map of fractional polarization along with a total intensity image and a map of the instrumental [547]-[1024] color \citep{smi01} is shown in Figure 3. The NW Arc is prominent in the total intensity image and is quite red in the color map. In the fractional polarization map this feature is clearly apparent as more weakly polarized than its surroundings. Arcs 1 and 2 are relatively blue in color and can be seen as moderately polarized features in the fractional polarization map. The highest fractional polarization, reaching 80\%, is found in the regions surrounding the NW Arc. The entire NE half of the nebula is faint and has relatively low polarization, varying between 10\% and 20\% on average. 

\section{Analysis} 

Polarimetry has proved a powerful tool for studying the geometry of extended circumstellar dust and gas, in particular bipolar systems. For example, \citet{sel92} were able to accurately fix the tilt of the bipolar outflow axis in the peculiar red giant OH 0739. \citet{sch99} were able to distinguish between different axisymmetric models for the ejecta surrounding Eta Carina. Polarimetry combined with theoretical modeling has been used to further refine the details of bipolar outflows and circumstellar disks \citep[e.g.][]{gle07, hof97}. In the visual images, VY CMa does not appear to have a mass loss nebula with any simple, underlying symmetry like a classic bipolar nebula. \citet{smi01} and \citet{hum05} argue that the circumstellar environment of VY CMa is dominated by complex, distinct arcs and features that were created at different epochs. Our goal is to use imaging polarimetry to locate each of these features in three dimensions with respect to the star, without reference to any model distribution.

The polarization of the nebulosity surrounding VY CMa is clearly due to scattering of light from the central star by dust in the circumstellar environment. The centro-symmetric pattern of position angles everywhere except directly East and the very high values of polarization we measure preclude other mechanisms, given that there is no significant ionized gas in the VY CMa system. For Rayleigh scattering, maximum polarization occurs at a scattering angle of exactly $90\degr$ in a plane perpendicular to our line of sight and containing the star (see Figure 4). For realistic circumstellar dust, the highest polarization will occur near $90\degr$, but depending on the scattering phase function of the dust, it will not be exactly at $90\degr$. The polarization will decrease as the scattering angle departs from this maximum in either direction.

For fractional polarizations less than the observed maximum, the dust must be either behind or in front of the plane of the sky where maximum polarization occurs, at a scattering angle away from optimum. This argument assumes there is only single scattering and no multiple scatters to depolarize the light. If the scattering is optically thin, we can compute two possible scattering angles from the observed fractional polarization, but we can not determine from the polarization alone whether the dust is at the larger or smaller scattering angle. This degeneracy is illustrated in Figure 4. In principle, light scattered from dust behind the plane of optimum scattering will experience more reddening exiting the nebula than light scattered from dust in front of that plane. This principle is also illustrated in Figure 4, and can provide a means of breaking the foreground/background degeneracy in locating the dust. There are other effects, such as pre-polarization of the light by scattering in very close to the star \citep[e.g.][]{min91} which we will ignore here.

This technique can not be expected to extract fine details of the geometry of the dust surrounding VY CMa, only large scale patterns. Any single line of sight through the nebulosity likely intersects dust at a range of scattering angles. The measured polarization represents a weighted average of the polarized and unpolarized intensities. Our technique places all of the emission at a single location based on the measured net polarization and color. Given these limitations, we will concentrate on regions that are clearly distinct in the polarization map, have very high polarization, or are very bright in total intensity. The details of our technique are given in the Appendix.

Critical to our technique is the assumption of single scattering, since multiple scatters will depolarize the light and cause our technique to locate the dust further behind or in front of the star than should be the case. We investigate this question and present the details of our calculations in the Appendix where we have computed the scattering optical depth at several locations in the nebula surrounding the star in the F658N filter. The results are listed in Table 2, where the nomenclature for the features is taken from Paper I. For the bright NW Arc, which is quite red in color, we obtain values for the scattering optical depth $\tau_s$ from 0.3 - 0.5, indicating that multiple scattering could be a contributing factor. For the bright clumps within $0.5\arcsec$ of the central star, we also find similar high estimates of $\tau_s$, indicating possible optically thick scattering. 

For most of the rest of the nebula, including Arc 1 and Arc 2, the very highly polarized regions $2-4 \arcsec$ West of the star, the S Arc and the low polarization regions N of the star, the computed values for $\tau_s$ range between 0.02 and 0.15. We will, therefore, assume optically thin scattering in those regions and use their polarization in the F658N filter and their [547]-[1024] colors to locate the scattering dust in three dimensions with respect to the star. Although we have observations in the F550M filter, in order to minimize the scattering optical depth we will concentrate on the F658N data. Since the NW Arc possibly suffers depolarization from multiple scattering, we will only be able to roughly determine the location of the material in this feature along our line of sight. We will find, however, that the polarization and color data suggest this arc may be two separate features.

\section{Discussion} 

Several previous studies of VY CMa report evidence for a bipolar outflow from the star \citep{kas98, shi03, yat93}. The derived geometry is not always consistent between these studies, however. \citet{shi03} observed SiO maser emission that shows a bipolar geometry and the emission comes entirely from the very inner $\sim 1\arcsec$ of the nebula. The only morphological resemblance between this SiO emission and the optical nebulosity is the difference in emission between the fainter optical nebulosity to the NE and the brighter nebulosity to the SW. Otherwise, the arcs and clumps seen in the optical nebulosity bear little resemblance to the limb brightened, hollow cone most commonly associated with classic bipolar outflows. The arcs, filaments and knots appear to be independent of any proposed bipolar geometry. Our polarimetry and images can not confirm nor rule out an underlying bipolar geometry to the mass-loss from VY CMa.

The most highly polarized regions in the nebula are to the West and SW of the NW Arc. Fractional polarizations range from 60-80\%. These polarizations place the dust close to the optimum scattering angle, about $100\degr$, which will be just behind the plane of the sky containing the star (see the Appendix). This material can not extend significantly in front or behind this location, or averaging along the line-of-sight will reduce the polarization below the high values we measure. There are no transverse velocity measurements for this diffuse nebulosity in Paper I for comparison with the polarimetry.

Arc 1 and Arc 2 are relatively blue features seen in total intensity to the South of the star. These features average about 40\% polarization and are surrounded by fainter, redder emission that is only 20-30\% polarized. Our techniuqe places these feautes in front of the star at an angle of about $35\degr$ to the plane of maximum polarization (about $25\degr$ to the star). This is in good agreement with the total space motions measured in Paper I, which suggest $33\degr$ for Arc 1 and $17\degr$ for Arc 2. The surrounding nebulosity is clearly behind these features, making them distinct geometric structures. If these features were just local density enhancements, the fractional polarization would not change across the visible edges in the intensity image. Any surrounding, but thinner, dust would scatter with the same geometry and produce the same fractional polarization, contrary to what is observed.

The NW arc is a prominent feature in the total intensity image and is extensively studied in \citet{hum05} and Paper I. It is clearly visible in the fractional polarization map (Figure 3) as a more weakly polarized region ($P \sim 40-60\%$) surrounded by highly polarized ($60 - 80\%$) nebulosity. In particular, the most Northwestern region of the feature is only 40\% polarized. This feature is also not likely to be a local density enhancement since it is so prominant in the polarization map. The brighter parts of the NW Arc are possibly optically thick, and this may contribute some to the the lower polarization. However, the fractional polarization across the feature is relatively uniform compared with the abrupt change in polarization at its edges. This indicates it is a distinct geometric feature, not just a local enhancement in dust density.

Using our technique, the red color and weaker polarization would place the NW Arc at an angle of $35-45\degr$ behind the star. This conclusion can be compared to the radial velocity and proper motion measurements discussed in \citet{hum05} and Paper I, subject to the caveat that the NW Arc is possibly optically thick (Table 2). The total space motion vectors (Paper I) for the more highly polarized inner (Eastern) edge of the NW Arc are consistent with a location for the scattering dust behind the plane of the sky at $\sim 25\degr$. There were no knots or clumps for use in measuring the proper motion of the outer (Western) edge of the NW Arc, but the radial velocities \citep{hum05} are all positive. This is consistent with placement behind the plane of the sky, but probably not as much as suggested by the polarimetry. Our polarimetry locates the Eastern edge at about $35\degr$ behind the plane of the sky and the Western edge at about $45\degr$. These results are reasonably consistent with the space motions.

In Paper I we suggested that the tip of the 'hook' at the end of the NW Arc, which appears to be bending back toward the star, may be nearer to us than the rest of the arc. The polarization rises from 40\% to 45\% along the hook to the tip and the color becomes slightly bluer. This would place the tip of the hook at a scattering angle closer to the plane of the sky than the bend in the hook, consistent with this suggestion.

Our analysis places the Eastern and Western portions of the NW Arc at different locations, suggesting it is more than one feature, or a relatively thin sheet seen partially edge on. This can be seen more clearly by taking a cut across the NW Arc, which is illustrated in Figure 5. The intensity, instrumental color and polarization along this cut are shown in Figure 6. Note that double peak in relative intensity that lines up with local maxima and minima in the color and polarization. On the Eastern side (left in Figure 6) the polarization has a local maximum and the color is slightly blue (with respect to an arbitrary zero). On the Western side the color is redder and the polarization lower. The NW Arc could be two features or perhaps a sheet seen mostly edge on with the Eastern side closer to the plane of the sky, with the Western edge further away. 

The S Arc is a rather faint feature in the total intensity map, but stands out in the fractional polarization map. It is slightly red in color, about 45\% polarized, and like Arcs 1 and 2, is surrounded by emission that is 20\% polarized. Our technique places the S Arc behind the star, about $40\degr$ from the plane of the sky. The spectrum across the S Arc shows two possible velocity components \citep{hum05}, so there may be some ambiguity as which radial velocity to associate with the S Arc material itself. In Paper I we used the velocity component that would place the S Arc in front of the plane of the sky containing the star, assuming radial motion out from the star. The other velocity component would have placed the S Arc slightly behind the star, closer to the location we have derived from the polarimetry. Although the location of the S Arc is somewhat ambiguous, the S Arc clearly stands out from the background in polarization and is very likely a distinct structure.

Most of the other features discussed in Paper I are very bright and within $1\arcsec$ of the star. These features tend to be relatively red are all likely to be optically thick in scattering (Table 1). The S Knot and the SW Clump are visible as separate features in the polarization map. They are polarized about $40\%$ and surrounded by fainter nebulosity that is about $20\%$ polarized. This suggests the S Knot and the SW Clump are distinct features, separate from the surrounding nebulosity. Due to the possibility of multiple scattering, we can not use the polarimetry to locate them along the line-of-sight. The W Arc has variations in polarization from $30 - 50\%$ and has no distinct morphology in the polarization map. This feature in the intensity image may be the line-of-sight composite of separate clumps of dust at differing scattering geometries with respect to the star. In Paper I we also suggested the W Arc may not be a single feature, but separate clumps along the line-of-sight.

\section{Summary}

We have used imaging polarimetry taken with the HST/ACS/HRC to explore the three dimensional structure of the circumstellar dust distribution around the red supergiant VY Canis Majoris. The fractional polarization ranges from a low of $10\%$ to a high of $80\%$, a relatively high value, even for reflection nebulosity. The polarization vectors on the sky show a strong centrosymmetric pattern in all directions except directly East of the central star.

In regions that are optically thin, and therefore likely have only single scattering, we used the fractional polarization and photometric color to locate the physical position of the dust along the line of sight. Most of the individual arc-like features and clumps appear to be separate geometric objects, and do not display any clear association with a simple bipolar outflow geometry. They are distinct both in the fractional polarization map and the intensity image. If these features were just local density enhancements, the fractional polarization would not change across strong brightness gradients. In all cases, the locations of these features are consistent with their observed proper motion and radial velocity from Paper I. 

Arcs 1 and 2 are well in front of the star and clearly separate from the surrounding nebulosity.  The bright NW Arc feature is possibly optically thick and can not be accurately located in space through polarimetry alone, but our polarimetry is consistent with the kinematics, which place the dust behind the plane of the sky containing the star. The diffuse material surrounding the NW Arc (as projected on the plane of the sky) is not associated with the Arc, but rather is in front of the NW Arc. The NW Arc does show evidence for either being two separate features, or a sheet of material seen partially edge-on. The edge closest to the star (the Eastern edge) is located closer to the plane of the sky than the redder, more weakly polarized Western edge, which is further away. 

\section{Appendix} 

We can estimate the scattering optical depth using the formula for optically thin scattering given in \citet{sel92}. Equation (6) in \cite{sel92} can be written as

\[
\tau _s  = \frac{{4\pi B_\lambda  \phi ^2 }}{{F_\lambda  \Phi (g_\lambda  ,\theta )\sin ^2 \theta }}
\]

where $B_\lambda$ is the observed surface brightness at angular offset $\phi$ from the central star with flux density $F_\lambda$ The phase function $\Phi (\theta)$ is given by

\[
\Phi (\theta ) = \frac{{4\pi S_{11} (\theta )}}{{\oint {S_{11} (\theta )d\omega } }} = \frac{{2S_{11} (\theta )}}{{\int_0^\pi  {S_{11} (\theta )\sin \theta d\theta } }}
\]

where $S_{11}$ is the familiar scattering matrix element \citep[p. 63]{boh83}, and we have assumed azimuthal symmetry in the scattering. Many authors use the Henevy-Greenstein \citep{hen41} formula for $\Phi$ with some assumptions about forward scattering. We will use direct Mie theory calculations instead.

Since the central star is mostly obscured and highly reddened at optical wavelengths along our particular line of sight, it would be difficult to determine the stellar flux density from visual observations alone. We can estimate $F_{658}$ for VY CMa by using the less reddened $H(1.65 \mu m)$ magnitude and assuming the typical $(R-H)$ color for an M5 supergiant. We can then compute an intrinsic R magnitude. We have chosen to use the $H$ magnitude as a compromise between the desire to use as long a wavelength as possible to minimize circumstellar extinction without including contamination by thermal dust emission, which is certainly a component at $K~(2.2 \mu m$, see discussion of the energy distribution in Paper I). 

Using $H=1.6$ from 2MASS \citep{skr06} and $(R-H) = 3.3$ \citep{joh67} we obtain an intrinsic $R$ magnitude of $R \sim 4.9$. Note that for the purpose of our rough estimates, we have ignored differences in the $R$ and F658N bandpasses. With a value for $F_{658}$ and our computed scattering phase function $\Phi(\theta)$, we can compute rough estimates of the scattering optical depth at any location in the nebula. We mention an important caveat, regions that are optically thick can have reduced surface brightness and yield artificially low computed values for $\tau_s$, although this can not be the case in regions of very high polarization. Very high polarization precludes multiple scattering in the relatively unrestricted geometry of the nebulosity surrounding VY CMa. Given the considerable uncertainties, we have set a conservative threshold for the transition from single scattering to multiple scattering at a computed value of $\tau_s = 0.15$. 

The very high polarization we observe in VY CMa precludes large, highly forward scattering grains in the ejecta, but there is no reason to expect the grains to be unusually tiny (Rayleigh) either. We will use dust parameters from models of dust shells around cool, luminous stars as a guide for choosing the dust parameters to use in our Mie calculations. In order to calculate the scattering and absorption efficiencies for astronomical silicates used in dust shell models, we chose optical constants $n = 1.8$ and $\kappa = 0.1$ \citep[e.g.][]{suh99}. The grain size commonly used in these dust shell models is $0.1 \mu m$. To avoid resonances in the Mie calculations from a single grain size, we use a simple size distribution evenly distributed from $0.05 - 0.15 \mu m$ (power law index of 0). The results of these calculations are shown in Figure 7. We find that the primary controlling parameter is the upper grain size limit. The power law index and the lower grain size limit have minor effect. The phase function we adopt in Figure 7 lies between pure Rayleigh scattering and scattering from large grains. We have slightly smoothed the computed polarization by averaging with a simple $10\degr$ box car. This is intended to somewhat compensate for the averaging over the line-of-sight. 

For any location in the nebula we can directly determine a pair of scattering angles from the observed polarization using our Mie calculations. Next we use the color to place the dust either in front, or behind the plane containing the star at the appropriate scattering angle. We could compute calibrated photometric colors across the nebula, but direct interpretation of those colors would depend on our admittedly rough dust grain model. For our purposes, we will consider the {\it instrumental} [547]-[1024] color for the nebulosity with the highest (80\%) polarization to be representative of dust at the optimum scattering angle (about $100\degr$). For instrumental colors redder than this value we will place the dust behind the plane containing the star. For bluer instrumental colors, we place the dust in front. This is illustrated in Figure 8 for selected features in the nebulosity of VY CMa.

We then took our fractional polarization image and a [547]-[1024] color image and computed the location of the dust in each pixel using MatLab. The MatLab analysis provided a full three dimensional distribution of the dust surrounding VY CMa assuming our technique was appropriate at every location. We found most of the fainter nebulosity located in between the brighter features seen in the total intensity image to be stretching back from the central star into the distance behind the plane of the sky containing the star. Bright regions that were highly polarized were located near the plane of the sky. Blue regions with modest polarization were located well in front of the star and moderately polarized redder regions were located behind the star, but not at the steep angle of the fainter, lower polarized nebulosity. Animations illustrating the three dimensional geometry distribution of the dust as computed by our technique are available online  \footnote{www.astro.umn.edu/$\sim$ahelton/research/VYCMa/}

\setcounter{figure}{0}
\clearpage
\begin{figure}
\plotone{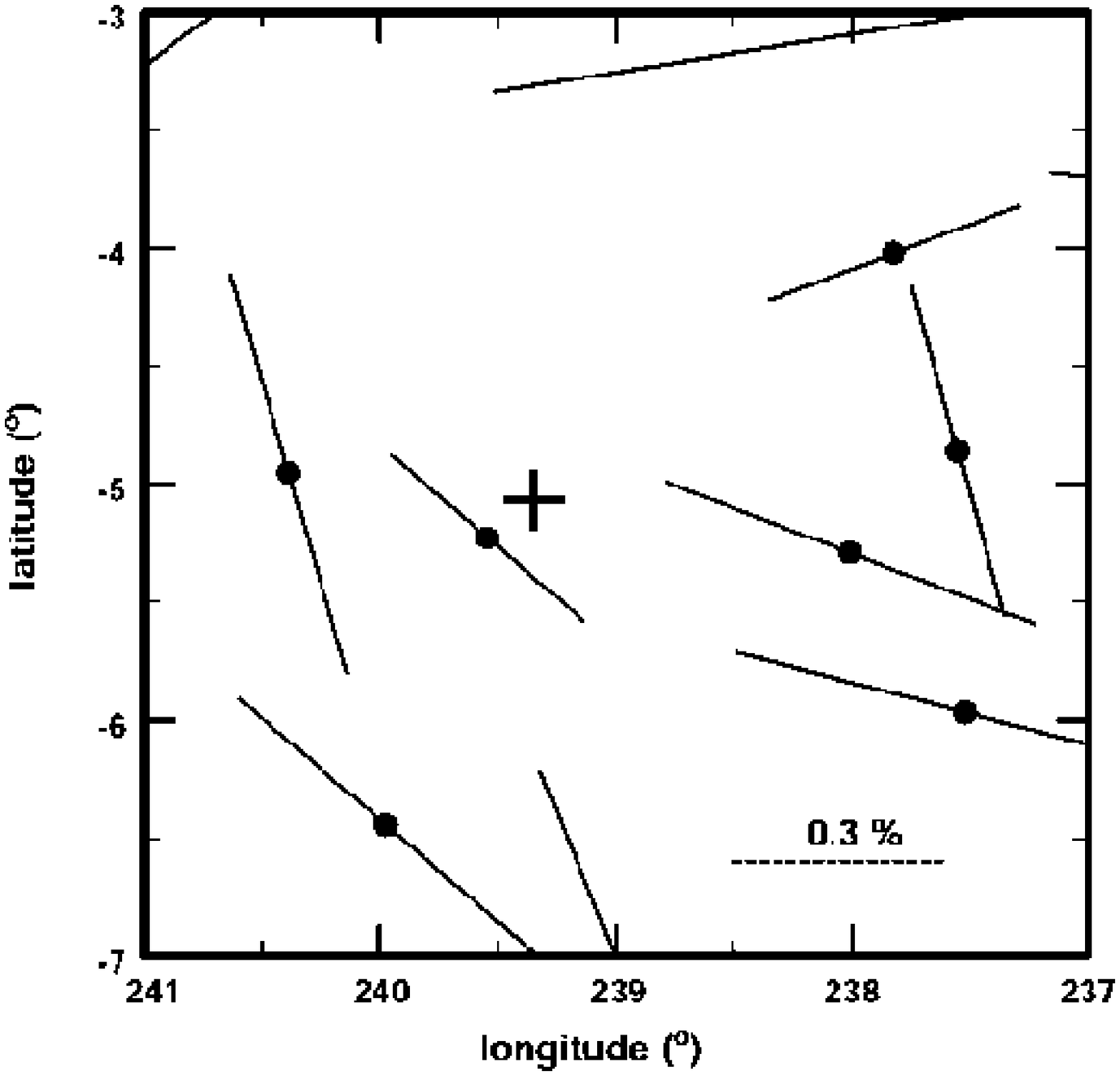}
\caption{Polarization vectors on the sky for stars in the vicinity of VY CMa, indicated by a + sign. Note the very low interstellar polarization in this direction.}
\end{figure}

\clearpage
\begin{figure}
\plotone{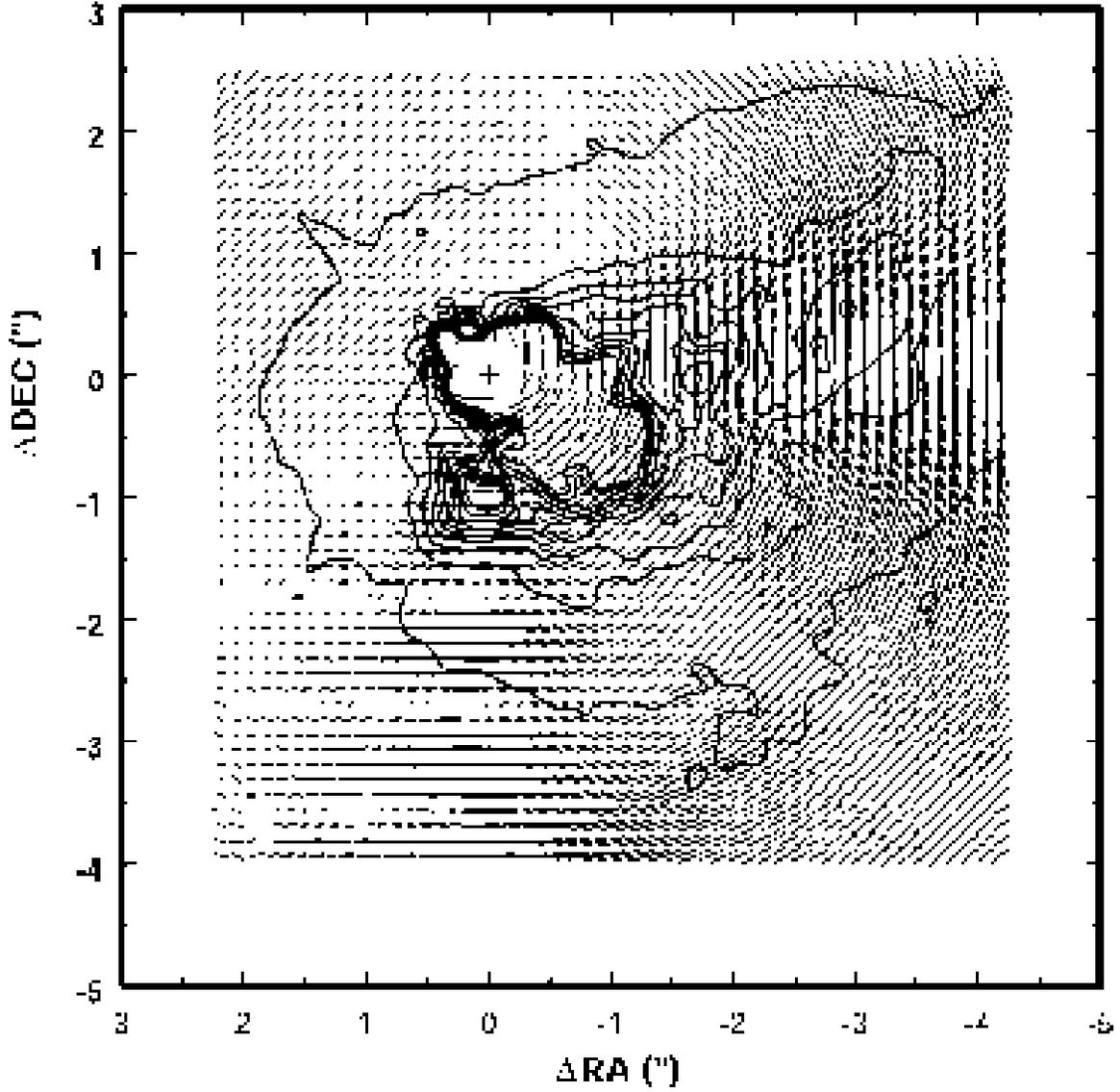}
\caption{Map of the polarization vectors for the nebulosity surrounding VyCMa. The longest vector is 80\%. The contours are total intensity in the F658N filer. }
\end{figure}

\clearpage
\begin{figure}
\plotone{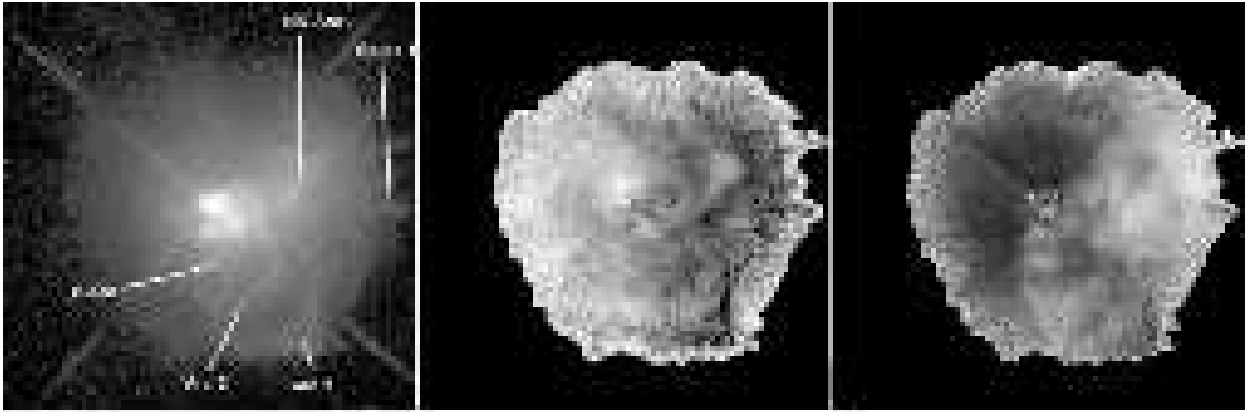}
\caption{Grey scale images of VY CMa at 658nm in total intensity (left), [547]-[1024] color (center, brighter = redder) and fractional polarization (right).}
\end{figure}

\clearpage
\begin{figure}
\plotone{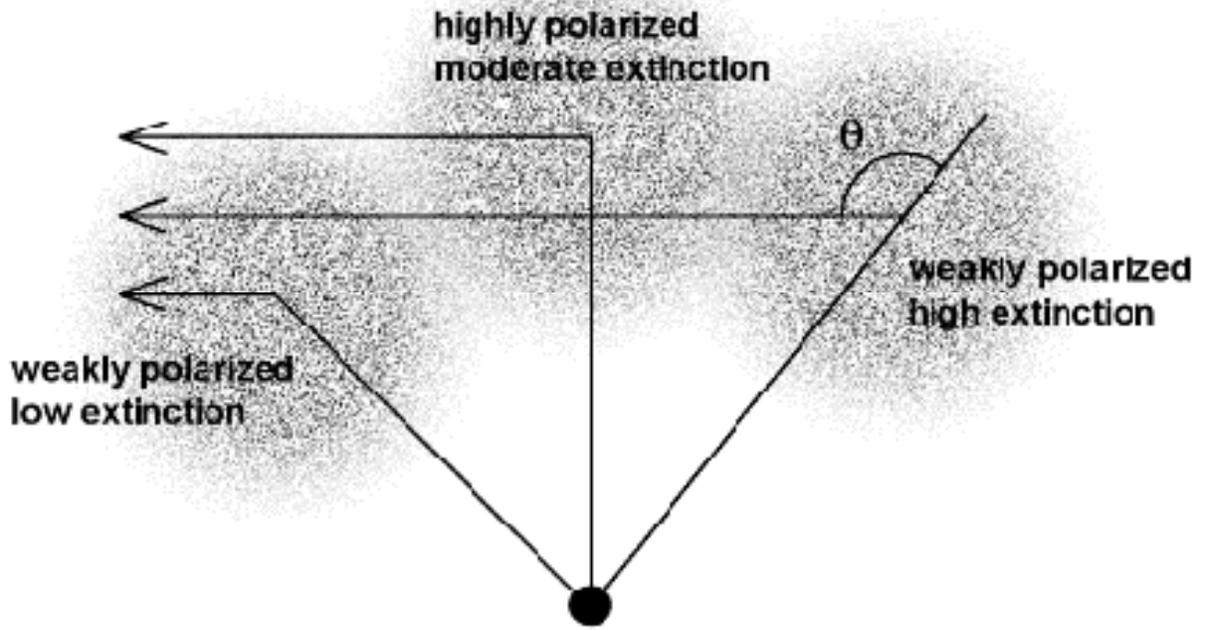}
\caption{Schematic drawing of the scattering of light from a central star at three different locations in the circumstellar environment. Note that the scattering angle $\theta$ is defined with respect to the line-of-sight from the star to the scattering location.}
\end{figure}

\clearpage
\begin{figure}
\plotone{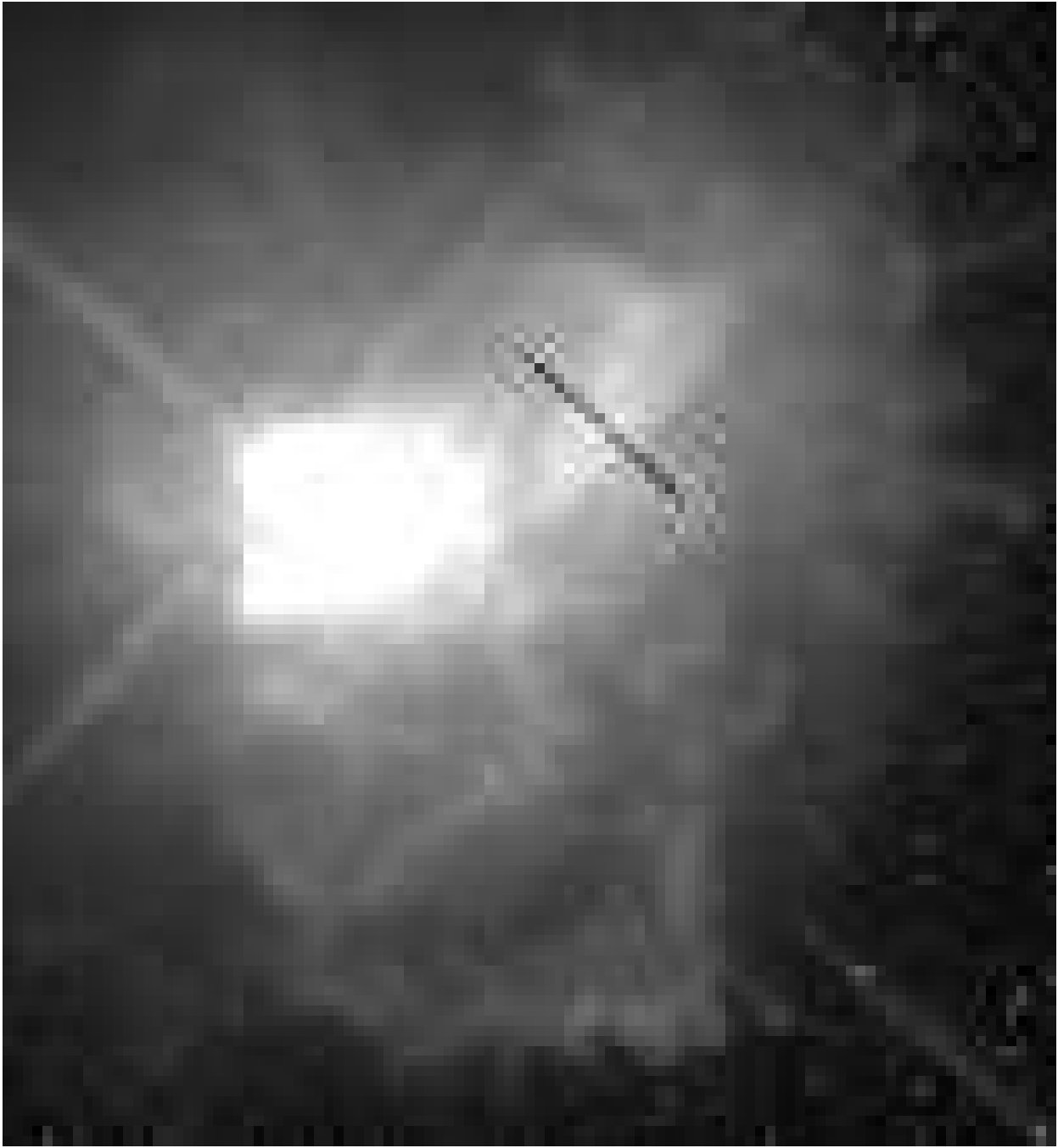}
\caption{Illustration of the location of a cut across the NW Arc in the image of VY CMa shown graphically in Figure 6. The direction of the cut is from the NE to the SW.}
\end{figure}

\clearpage
\begin{figure}
\plotone{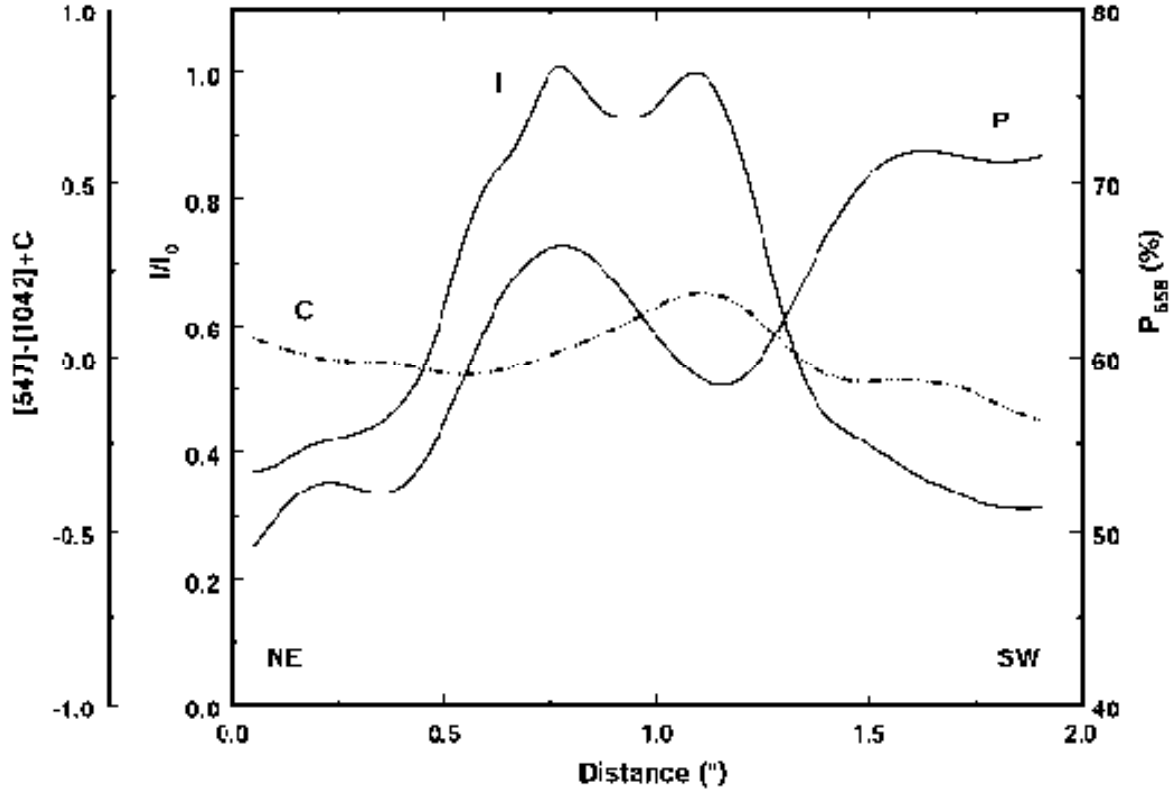}
\caption{Graphical representation of the cut across the NW Arc illustrated in Figure 6 for the quantities relative intensity (I), instrumental [547]-[1024] color (C) and fractional polarization (P). Note that there are two peaks in intensity that correspond to reversals in polarization and color. The inner (left) edge of the NW Arc is bluer and more highly polarized, placing it closer to the plane of the sky than the outer (right) edge of the feature which is redder and more weakly polarized.}
\end{figure}

\clearpage
\begin{figure}
\plotone{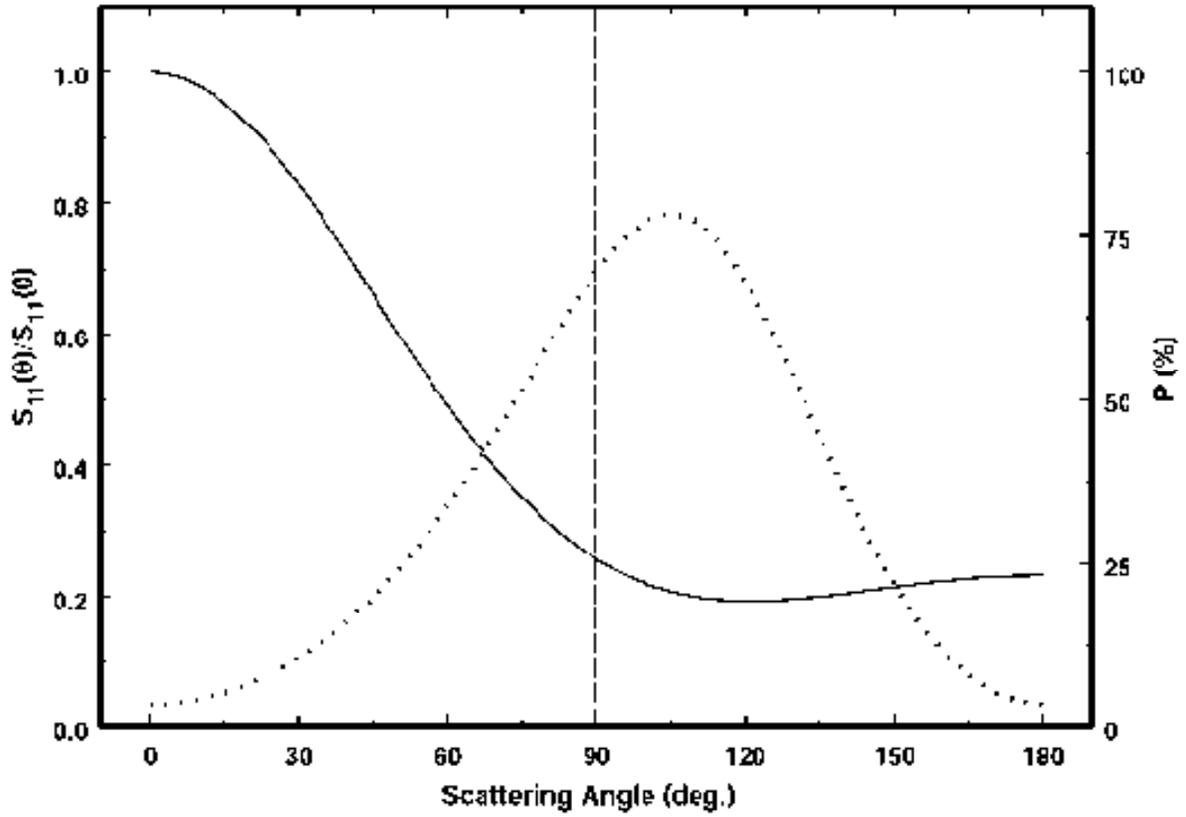}
\caption{Computed scattering phase function (solid line) and fractional polarization (dashed line) for typical optical parameters for dust surrounding luminous, cool, oxygen rich stars.}
\end{figure}

\clearpage
\begin{figure}
\plotone{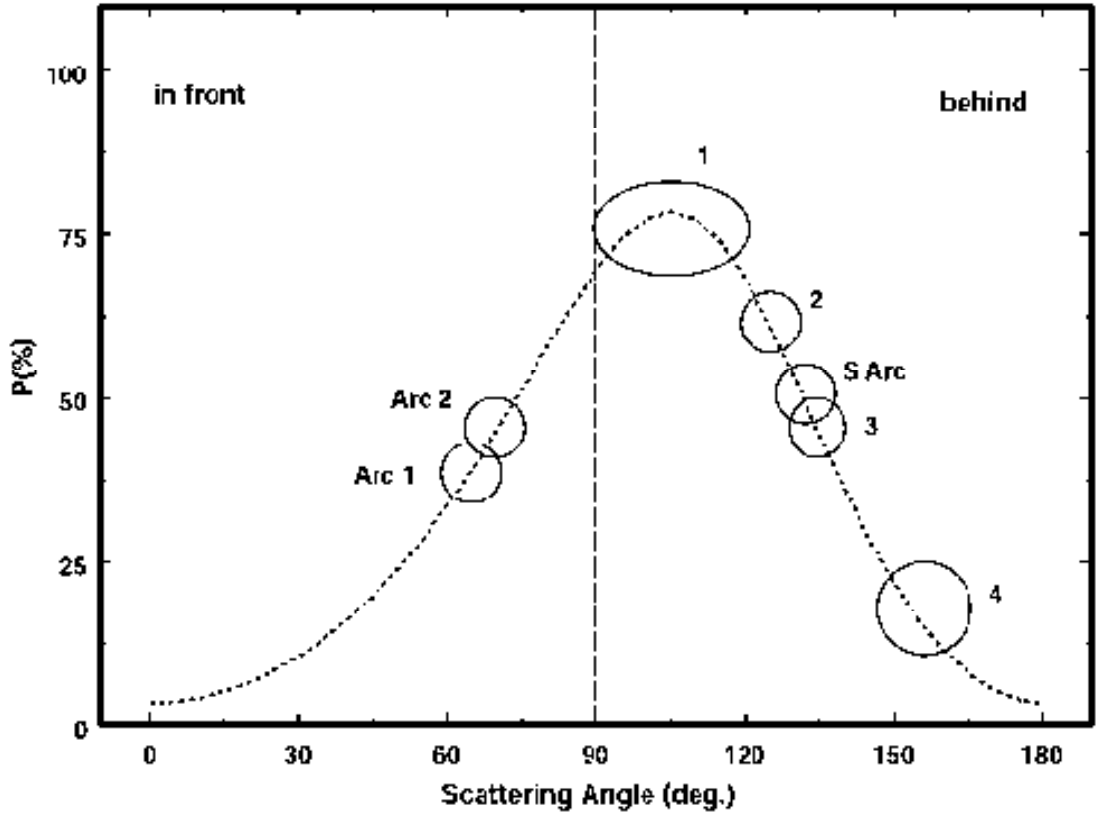}
\caption{Location of selected features in scattering angle based on their fractional polarization and color. The vertical dotted line represents the plane of the sky containing the central star. Small scattering angles (forward scattering) lie in front of the plane containing the star. Large scattering angles (back scattering) line behind this plane. The plane of maximum polarization does not correspond exactly to $90\degr$ scattering (the plane of the sky), but lies at $\sim 100\degr$, about $10\degr$ behind the plane of the sky. The numbered regions correspond to: 1) the highly polarized nebulosity to the West and SW of the NW Arc, 2) the inner (Eastern) edge of the NW Arc, 3) the outer (Western) edge of the NW Arc, and 4) the redder, faint nebulosity filling in between Arc 1 and Arc 2.}
\end{figure}

\begin{deluxetable}{cccc}
\tablenum{1}
\tablewidth{0pt}
\tablecaption{Offset Star Polarization}
\tablehead{
\colhead{Column Offset} & \colhead{Row Offset} & \colhead{$P_{X}(\%)$} & \colhead{$P_{Y}(\%)$}
}
\startdata

0 & 0 & -6.80 & 0.37 \\
153 & -141 & -6.58 & -.72 \\
-137 & 128 & -6.28 & -.69 \\

\enddata
\end{deluxetable}

\begin{deluxetable}{ccc}
\tablenum{2}
\tablewidth{0pt}
\tablecaption{Minimum Scattering Optical Depth}
\tablehead{
\colhead{Location} & \colhead{$\tau_s$} & \colhead{P(\%)} 
}
\startdata

Arc 1 A & 0.08 & 35 \\
Arc 1 B & 0.11 & 30 \\
Arc 2 B & 0.09 & 40 \\
S Arc A & 0.09 & 50 \\
Spike 1 A & 0.02 & 45 \\
SE Knot & 0.04 & 15 \\
NW Arc B & 0.5 & 50 \\
SW Clump & 0.38 & 45 \\
S Knot & 0.20 & 45 \\

\enddata
\end{deluxetable}


\begin{thebibliography}{}

\bibitem[Bohren \& Huffman (1983)]{boh83} Bohren, C.~F. \& Huffman, D.~R.\ 1983, {\it Absorption and Scattering of Light by Small Particles} (Wiley-Interscience, New York), p. 63.

\bibitem[Bowers et al.(1983)]{bow83} Bowers, P.~F., Johnston, 
K.~J., \& Spencer, J.~H.\ 1983, \apj, 274, 733

\bibitem[Danchi et al.(1994)]{dan94} Danchi, W.~C., Bester, 
M., Degiacomi, C.~G., Greenhill, L.~J., \& Townes, C.~H.\ 1994, \aj, 107, 
1469

\bibitem[de Jager(1998)]{dej98} de Jager, C.\ 1998, \aapr, 8, 
145

\bibitem[Heiles(2000)]{hei00} Heiles, C.\ 2000, \aj, 119, 923

\bibitem[Henyey \& Greenstein(1941)]{hen41} Henyey, L.~G., \& 
Greenstein, J.~L.\ 1941, \apj, 93, 70

\bibitem[Herbig(1972)]{her72} Herbig, G.~H.\ 1972, \apj, 172, 
375

\bibitem[Hoffman et al.(1997)]{hof97} Hoffman, J.~L., 
Whitney, B.~A., Wood, K., \& Nordsieck, K.~H.\ 1997, Bulletin of the 
American Astronomical Society, 29, 1279 

\bibitem[Humphreys \& Davidson(1994)]{HD} Humphreys, 
R.~M., \& Davidson, K.\ 1994, \pasp, 106, 1025

\bibitem[Humphreys et al.(2005)]{hum05} Humphreys, R.~M., 
Davidson, K., Ruch, G., \& Wallerstein, G.\ 2005, \aj, 129, 492

\bibitem[Humphreys et al.(2007)]{hum07} Humphreys, R.~M., Helton, L.~A., \& Jones, T.~J.\ 2007,
\aj, submitted

\bibitem[Johnson(1967)]{joh67} Johnson, H.~L.\ 1967, \apj, 
149, 345

\bibitem[Jones \& Gehrz(2000)]{jon00} Jones, T.~J., \& Gehrz, 
R.~D.\ 2000, Icarus, 143, 338 

\bibitem[Jones(1997)]{jon97} Jones, T.~J.\ 1997, \aj, 114, 
1393 

\bibitem[Kastner \& Weintraub(1998)]{kas98} Kastner, J.~H., 
\& Weintraub, D.~A.\ 1998, \aj, 115, 1592

\bibitem[Kozhurina-Platais \& Biretta (2005)]{koz05} Kozhurina-Platais \& V., Biretta, J.\ 2005, {\it{Instrument Science Report}} ACS 2005-10

\bibitem[Lada \& Reid(1978)]{lad78} Lada, C.~J., \& Reid, 
M.~J.\ 1978, \apj, 219, 95 

\bibitem[Lowe \& Gledhill(2007)]{gle07} Lowe, K.~T.~E., \& 
Gledhill, T.~M.\ 2007, \mnras, 374, 176

\bibitem[Marvel(1997)]{mar97} Marvel, K.\ 1997, \pasp, 109, 
1286 

\bibitem[Minchin et al.(1991)]{min91} Minchin, N.~R., Hough, 
J.~H., McCall, A., Aspin, C., Yamashita, T., \& Burton, M.~G.\ 1991, 
\mnras, 249, 707

\bibitem[Morris \& Bowers(1980)]{mor80} Morris, M., \& 
Bowers, P.~F.\ 1980, \aj, 85, 724

\bibitem[Perlman et al.(2006)]{per06} Perlman, E.~S., et al.\ 
2006, \apj, 651, 735 

\bibitem[Richards et al.(1998)]{ric98} Richards, A.~M.~S., 
Yates, J.~A., \& Cohen, R.~J.\ 1998, \mnras, 299, 319

\bibitem[Schulte-Ladbeck et al.(1999)]{sch99} 
Schulte-Ladbeck, R.~E., Pasquali, A., Clampin, M., Nota, A., Hillier, 
D.~J., \& Lupie, O.~L.\ 1999, \aj, 118, 1320

\bibitem[Sellgren et al.(1992)]{sel92} Sellgren, K., Werner, 
M.~W., \& Dinerstein, H.~L.\ 1992, \apj, 400, 238

\bibitem[Shinnaga et al.(2003)]{shi03} Shinnaga, H., 
Claussen, M.~J., Lim, J., Dinh-van-Trung, \& Tsuboi, M.\ 2003, ASSL 
Vol.~283: Mass-Losing Pulsating Stars and their Circumstellar Matter, 393

\bibitem[Skrutskie et al.(2006)]{skr06} Skrutskie, M.~F., et 
al.\ 2006, \aj, 131, 1163

\bibitem[Smith et al.(2001)]{smi01} Smith, N., Humphreys, 
R.~M., Davidson, K., Gehrz, R.~D., Schuster, M.~T., \& Krautter, J.\ 2001, 
\aj, 121, 1111 

\bibitem[Suh(1999)]{suh99} Suh, K.-W.\ 1999, \mnras, 304, 389

\bibitem[Yates(1993)]{yat93} Yates, J.~A.\ 1993, LNP 
Vol.~412: Astrophysical Masers, 412, 415

\end{thebibliography}
\end{document}